\documentclass{article}
\usepackage{spconf,amsmath,graphicx}
\usepackage{booktabs, multirow, amssymb, anyfontsize, glossaries, soul, color}

\usepackage{subcaption, balance}

\title{Video Compression Beyond VVC: Quantitative Analysis of Intra Coding Tools in Enhanced Compression Model (ECM)}
\name{Mohsen Abdoli$^{\star}$ \qquad Ramin G. Youvalari$^{\dagger}$ \qquad Karam Naser$^{\ddagger}$ \qquad Kevin Reuz\'{e}$^{\ddagger}$ \qquad Fabrice Le L\'{e}annec$^{\ddagger}$}

\address{$^{\star}$ Xiaomi Technologies, Rennes, France \\
$^{\dagger}$ Xiaomi Technologies, Tampere, Finland  \\
$^{\ddagger}$ InterDigital, Rennes, France 
}

\begin{document}
%
\maketitle
%


\begin{abstract}
A quantitative analysis of post-VVC luma and chroma intra tools is presented, focusing on their statistical behaviors, in terms of block selection rate under different conditions. The aim is to provide insights to the standardization community, offering a clearer understanding of interactions between tools and assisting in the design of an optimal combination of these novel tools when the JVET enters the standardization phase. Specifically, this paper examines the selection rate of intra tools as function of 1) the version of the ECM, 2) video resolution, and 3) video bitrate. Additionally, tests have been conducted on sequences beyond the JVET CTC database. The statistics show several trends and interactions, with various strength, between coding tools of both luma and chroma.
\end{abstract}
\section{Introduction}
\label{sec:intro}
While the Versatile Video Coding (VVC) standard is on the verge of industrial adoption \cite{bross2021overview, hamidouche2022versatile}, the Joint Video Experts Team (JVET) is actively engaged in the development of the next-generation video codec to succeed VVC. This project focuses on the identification of coding tools that can significantly enhance the compression efficiency of VVC \cite{li2024ahg7}. The selected tools are integrated into a reference software model known as the Enhanced Compression Model (ECM) and are being continuously improved to find the optimal configuration \cite{coban2024ecm}. As of the writing of this paper, the JVET is in \textit{exploration} phase which is typically followed by a formal \textit{standardization} phase, in which definitive decisions will be made to specify the normative functionality of best proposals. 

Due to the incremental adoption of the proposed tools across multiple JVET meetings, the intricate interactions between their technical aspects can be challenging to measure. This is due to the complexity of the reference software as well as growing number of tools that make numerous combinations to study, which is highly resource demanding. However, in order to develop a reliable next-generation codec with high performance in real-life scenarios (which increases its chance of industrial adoption), it is essential for the standardization community to overcome this difficulty and understand the interactions of different compression tools \cite{liu2022statistical, abdoli2024stats}.

A quantitative analysis of the tools adopted up to the version 11.0 of the ECM is provided to fill the above gap and offer insights into individual tools developed for video compression beyond VVC. To this end, this paper studies statistics (\textit{i.e.} the selection rate) of major intra coding tools that, as of the time of this publication, are under development within the ECM. These tools are categorized into luma intra coding tools and chroma intra coding tools. The rest of this paper is organized as follows: Section~\ref{sec:tools} briefly introduces the set of selected luma and chroma intra tools in ECM. Section~\ref{sec:stats} provides quantitative figures showing the statistical behaviour of the selected tools, and Section~\ref{sec:conclusion} concludes the paper.



\section{Intra coding tools in ECM}
\label{sec:tools}

Tools presented in this section are a high-level subset of intra coding tools that are currently used for natural content coding with ECM. It is important to note that it was the authors' wish to have, for clarity, mutually excluding tools for a sum of use of 100\%. In particular, Intra Sub Partition (ISP) and Multi Reference Line (MRL) can only be used in combination of other tools and were, therefore, not included \cite{coban2024ecm}. Furthermore, while Neural Network based Video Coding (NNVC) \cite{galpin2024nnvc} holds significant potential, it is deemed beyond the scope of this paper and, therefore, is not addressed herein.

\subsection{Luma coding tools}

\textit{Template-based Intra Mode Derivation (TIMD)}: For each intra prediction mode in the Most Probable Modes (MPM) list, the Sum of Absolute Transformed Differences (SATD) between the prediction and reconstructed neighbor samples of a template is calculated. The intra prediction mode with the minimum SATD is selected as the TIMD mode and used for the prediction of the current block \cite{cao2021timd}.

\textit{Decoder-side Intra Mode Derivation (DIMD)}: When DIMD is applied, five intra modes are derived from the reconstructed neighbor samples, and are combined with the planar mode predictor with the weights derived from a Histogram of Gradients (HoG). The HoG computation is carried out by applying horizontal and vertical Sobel filters on pixels in a template of width 3 around the block \cite{abdoli2020dimd}.

\begin{table*}[t] 
  \centering
  \begin{tabular}{c|ccccccccccc}
    
    & v1.0 & v2.0 & v3.0 & v4.0 & v5.0 & v6.0 & v7.0 & v8.0 & v9.0 & v10.0 & v11.0 \\
    \hline
    \hline
    Y & -5.0 \%&	-5.9 \%&	-6.1 \%&	-6.5 \%&	-6.8 \%&	-8.2 \%&	-9.1 \%&	-9.9 \%&	-11.6 \%&	-12.5 \%&	-12.8 \% \\
    \hline
    Cb & -7.3 \%&	-9.2 \%&	-11.7 \%&	-12.4 \%&	-14.1 \%&	-16.5 \%&	-18.0 \%&	-19.2 \%&	-22.7 \%&	-23.5 \%&	-23.7 \% \\
    \hline
    Cr & -7.4 \%&	-9.6 \%&	-12.9 \%&	-13.6 \%&	-15.3 \%&	-17.6 \%&	-19.0 \%&	-20.2 \%&	-23.7 \%&	-24.6 \%&	-24.8 \% \\
    
  \end{tabular}
  \caption{BD-Rate performance of ECM versions over VTM in AI configuration on Y, Cb and Cr components. 
  }
  \label{tab:mytable}
\end{table*}

\textit{Spatial Geometric Partition Mode (SGPM)}: SGPM is an intra mode that resembles the inter coding tool of GPM \cite{coban2024ecm}, where two prediction parts are generated from the intra process. In this mode, a candidate list is built with each entry containing one partition split and two intra modes \cite{wang2022sgpm}.

\textit{Template-based Multiple Reference Line (TMRL)}: An extended reference line list of 5 candidates and an extended MPM list 10 candidates are constructed. The area in reference line 0 is used for template matching. The SAD costs over the template area are calculated between the predictions (generated by 50 combinations) and the reconstructions. The 20 combinations with the least SAD cost are selected in an ascending order to form the TMRL candidate list, and a combination is signaled to the decoder \cite{xu2022tmrl}. 

\textit{Matrix-based Intra Prediction (MIP)}: For a block of size $W\times H$, as input, MIP takes one line of $H$ ($W$) reconstructed neighbouring boundary samples from left (above) of the block reconstructed neighbouring boundary samples. The generation of the prediction signal is based on three steps, which are averaging of the neighboring samples, matrix vector multiplication and linear interpolation \cite{pfaff2020data}.

\textit{Intra with Template Matching (IntraTmp)} and Intra Block Copy (IBC): Both are intra modes initially designed for screen content coding and later improved for natural contents, where the prediction is generated by copying reconstructed blocks specified by Block Vectors (BV) poiting to a reference block in the same frame. The main difference between IntraTmp \cite{naser2022itmp, wang2022itmp, youvalari2023filtered} and IBC \cite{chen2023ibc} is that BVs are derived instead of being signaled. That is to say, the decoder employs template matching to find the best block vectors instead of parsing their corresponding syntax element. 



\textit{Regular luma Intra Prediction Mode (IPM)}: In ECM, as in VVC, 65 angular IPMs with block shape-adaptive directions and 4-tap interpolation filters are supported as well as the DC and Planar modes. A list of Most Probable Modes (MPM) is created based on neighboring information and a lower signaling cost is given to IPMs in the MPM list \cite{pfaff2021intra}.

\subsection{Chroma coding tools}

\textit{DIMD}: This is similar to luma DIMD. The reconstructed templates of both luma and chroma components are analyzed to obtain a histogram of oriented gradients. The intra mode direction corresponding to the maximum histogram value is used for predicting the current block \cite{lainema2022dimd, abdoli2020gradient}.

\textit{Cross-Component Linear Model (CCLM)}: In VVC, the CCLM tool employs a linear model with two parameters to predict chroma samples based on luma samples. The model parameters are determined using neighboring samples of the block. In ECM, this functionality has been enhanced to incorporate additional reference samples from the block's surroundings. Furthermore, ECM introduces a multi-model variant, which involves deriving two linear models for block prediction. This is achieved by categorizing reference samples into two classes, using the average luma reference samples as the classification parameter \cite{ghaznavi2020joint}.

\textit{Convolutional Cross-Component Model (CCCM)}: Similar to CCLM, the CCCM tool also predicts chroma samples from luma samples. However, CCCM takes a different approach by utilizing reconstructed luma and chroma samples to establish a two-dimensional convolutional model, incorporating both a bias term and a nonlinear term. It also uses up to six reference lines for calculating the parameters \cite{astola2022cccm}.

\textit{Gradient-based Linear Model (GLM)}: In contrast to CCCM, this mode uses additionally the gradient of luma samples to derive a linear model. Specifically, two GLM modes are defined, one using the gradient only and the other uses the luma value and its gradient. The linear model is derived from the reconstructed template along with bias value.\cite{astola2022glm}

\textit{Cross Component Prediction (CCP) merge}: In this mode, a list of previously used cross-component prediction (CCP) modes is constructed, named CCP merge list. A list index is signaled to indicate which mode is used. The decoder performs the corresponding cross-component model using the stored cross-component parameters \cite{zhang2023ccp}.

\textit{Direct Block Vector (DBV)}: Similar to IBC and IntraTMP, this mode uses block vectors to generate the chroma prediction signals. Instead of signaling/deriving the block vectors, the ones from the collocated luma blocks are reused \cite{huo2022dbv}.

\textit{Regular chroma IPM} and Direct Mode (DM): Here, the regular mode refers to the remaining directional, DC and planar modes (similar to luma regular IPMs) \cite{pfaff2021intra}. Moreover, the Direct Mode (DM) is also used for a chroma block that simply uses the intra mode of the co-located luma intra block.

In the statistics of this paper, luma intra coding tools are categorized into three groups of 1) IPM-based (\textit{i.e.} regular IPM, DIMD, TIMD, SGPM and TMRL), 2) BV-based (\textit{i.e.} IBC and IntraTmp) and 3) others (\textit{i.e.} MIP). Similarly, chroma intra coding tools are categorized into 1) IPM-based (\textit{i.e.} regular IPM, DM and DIMD), 2) cross-component (\textit{i.e.} CCLM, CCCM, GLM and CCP merge) and 3) other (\textit{i.e.} DBV).


\begin{figure*}
    \centering
    \includegraphics[width=1\textwidth]{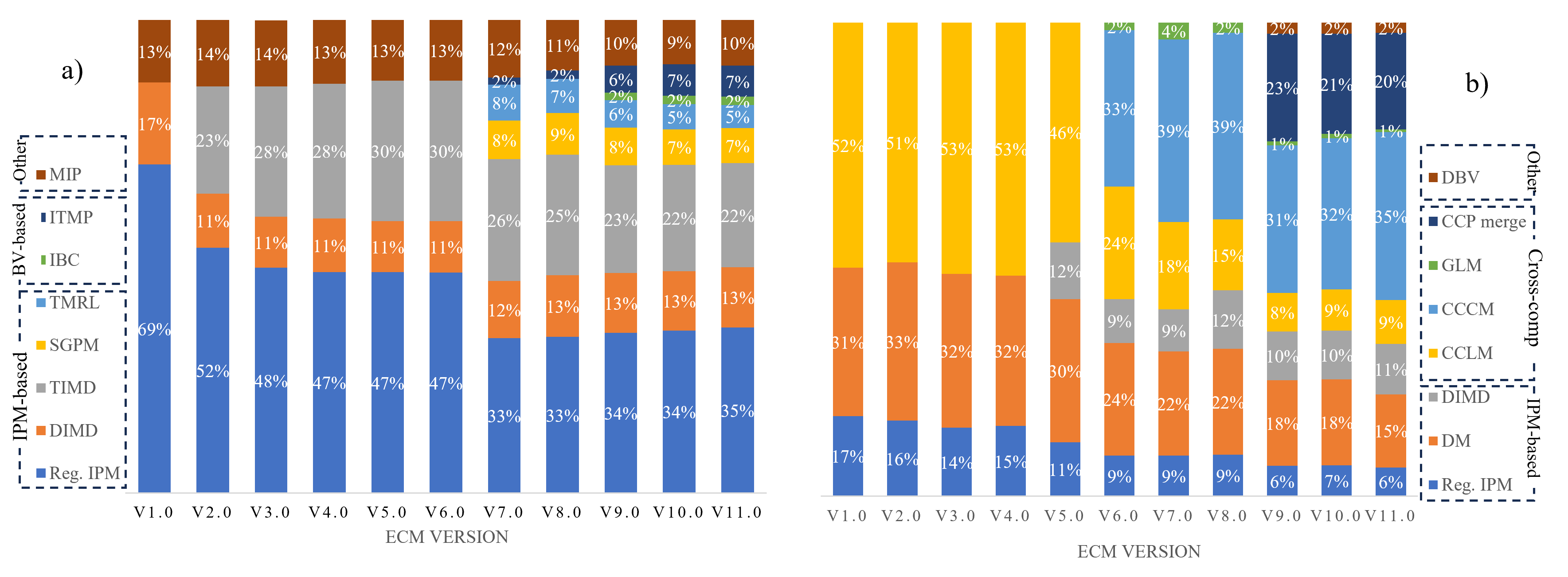}
    \caption{Evolution of selection rate of luma (a) and chroma (b) intra tools in different ECM versions.}
    \label{fig:versions}
\end{figure*}

\section{Quantitative analysis}
\label{sec:stats}
In this paper, the focus is on the selection rate of the coding tools. Due to the limitations of space, neither of the presented analysis are categorized based on the block size. It is also important to note that even though the selection rate of a tool has a high correlation with its BD-rate gain, it is also possible that a tool with lower selection rate brings higher bitrate saving gain compared to another tool with higher selection rate. For further information on per-tool performance of ECM, one can refer to \cite{li2024ahg7}.

Table~\ref{tab:mytable} shows the BD-Rate performance of ECM versions over VTM in All-Intra (AI) configurations. The latest ECM-11.0 software provides -12.8 \%, -23.7 \%, -24.8 \% coding gain over VTM in luma and chroma components, respectively. 
As can be seen, throughout versions of ECM, compression performance in all three components increase consistently. 

Fig.~\ref{fig:versions} illustrates the selection rate statistics of intra tools on different versions of ECM for luma and chroma. 
In general, adoption of a new tool might or might not impact the behaviour of other tools. This is due to the fact that some times coding tools exploit the same type of redundancies, hence their joint performance can be impacted when both enabled. One general trend that can be observed both in luma and chroma is that the selection rate of regular IPM modes have been significantly decreased throughout ECM version and following the adoption of new intra coding tools. This may be due to the fact that regular IPM is the last mode to signal in the normative list of coding tools. Therefore, addition of a new tool that is typically signaled before regular IPMs, would make it even more expensive in terms of rate, hence, less likely to be selected for a block. Moreover, new tools tend to be more advanced (\textit{e.g.} data-driven, requiring decoder side parameter derivation) and outperform regular IPM that has not fundamentally changed much in many years.

\begin{figure}
    \centering
    \includegraphics[width=.5\textwidth]{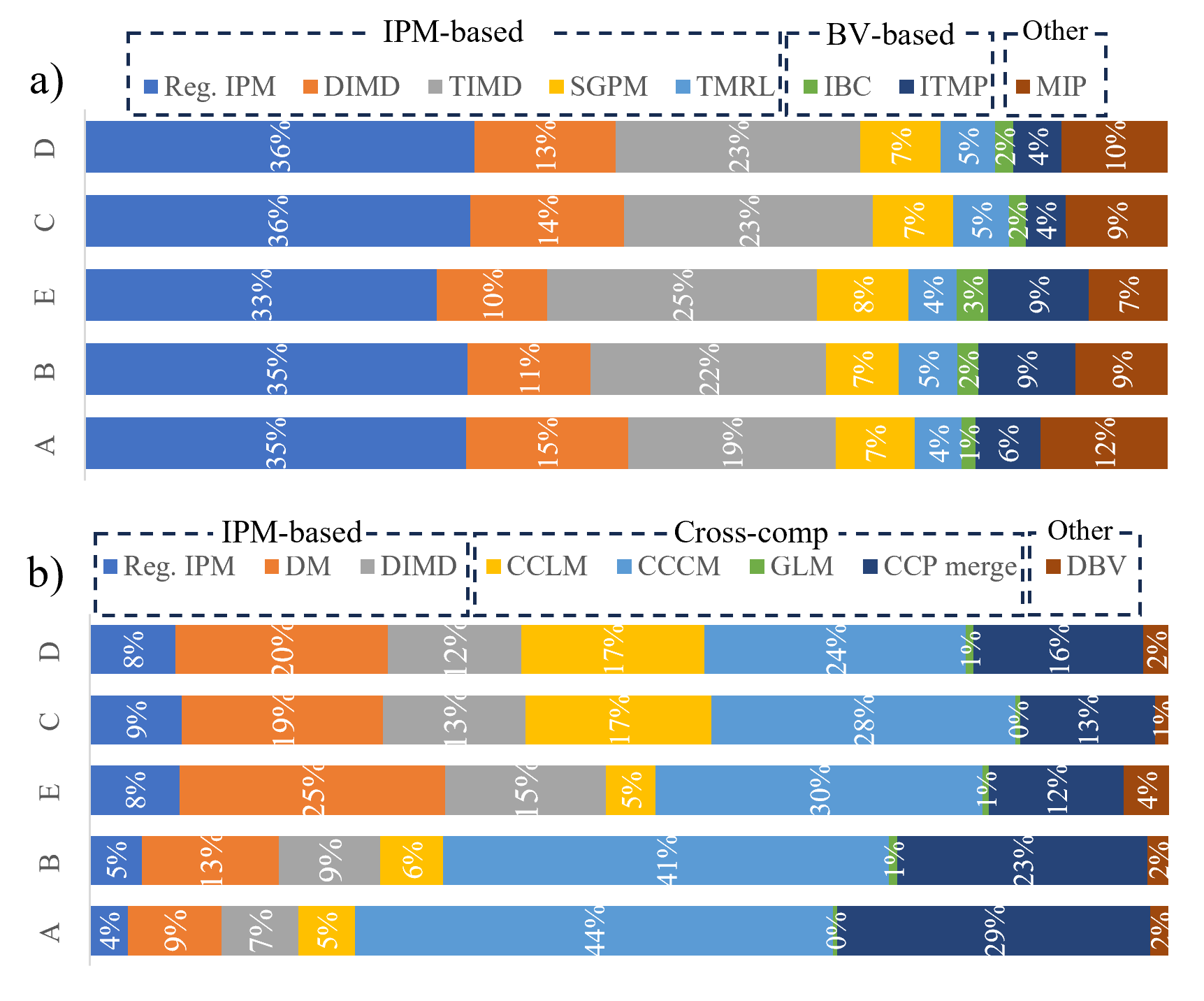}
    \caption{Selection rate of luma (a) and chroma (b) coding tools on CTC classes (ECM-11.0).}
    \label{fig:class}
\end{figure}

In luma, three coding tools of MIP, TIMD and DIMD seem to be the most stable tools. Roughly, this means that the adoption of new coding tools did not have major impact on their performance.
 Specifically, the statistics shows that unlike regular IPMs, coding of MIP blocks cannot easily be replaced by new tools, proving that it is targeting a particular type of spatial texture correlation. Another interesting observation is that, despite relatively low selection rate of IBC (around 2\%), its usage for natural content accounts for about 1\% additional bitrate saving \cite{chen2023ibc}.
In chroma coding, CCCM and DIMD are the most stable coding tools that have not been significantly impacted by adoption of other tools. 

Fig~\ref{fig:class} show the selection statistics of luma and chroma tools on different classes of the JVET CTC database \cite{Karczewicz2023ctc}. In luma, the selection rate of tools are mostly stable through different classes of resolution. In chroma, three major trends can be observed. First, the CCCM selection rate tends to increase with the video resolution. Second, CCP merge also has a higher selection rate in large resolutions. Finally, the DM mode is selected less often in high resolutions.

\begin{figure}
    \centering
    \includegraphics[width=.5\textwidth]{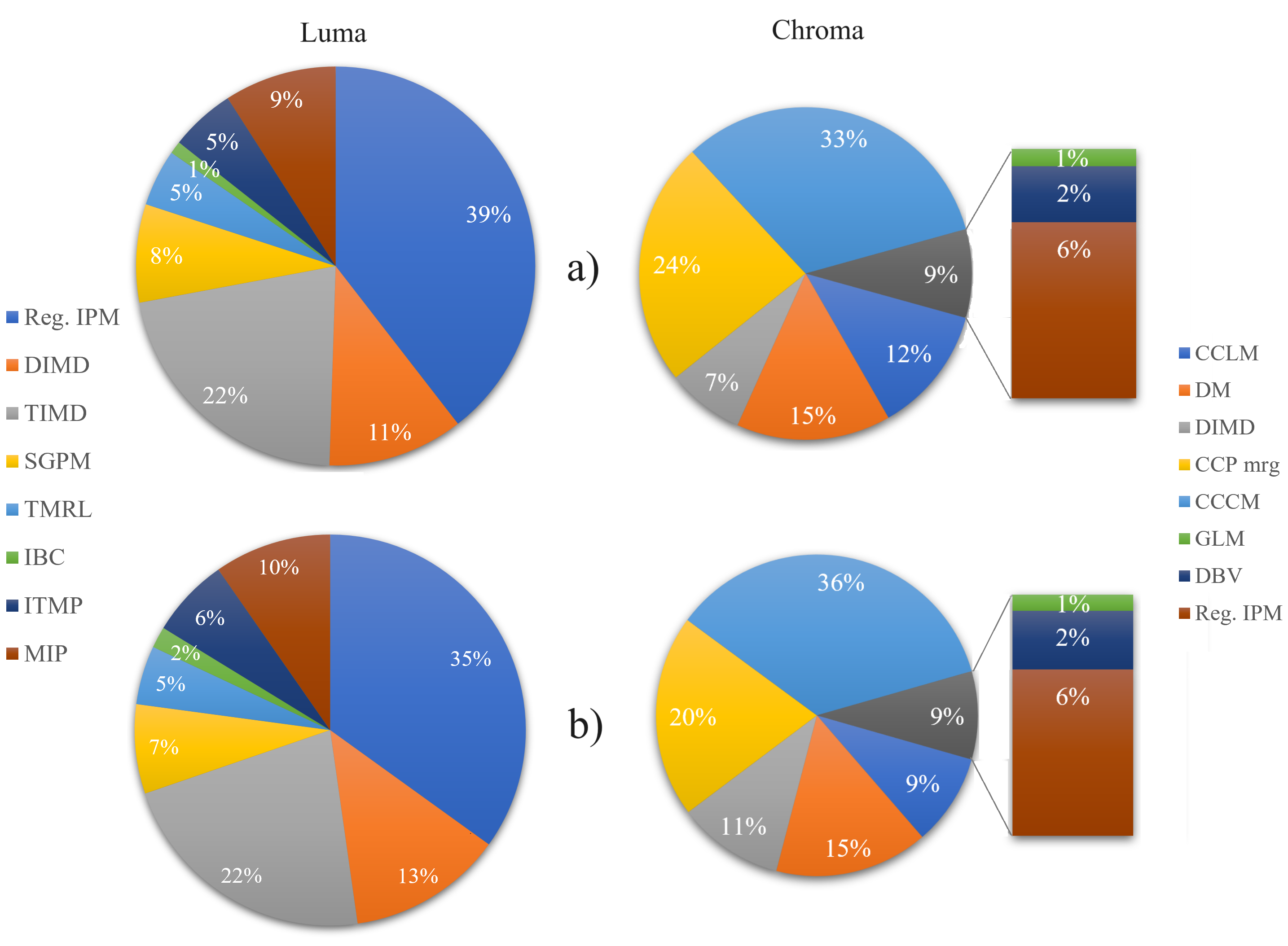}
    \caption{Selection rate of coding tools (ECM-11.0) on BVI-DVC (a) vs. CTC (b) for luma (left) and chroma (right).}
    \label{fig:bvi-ctc}
\end{figure}

Fig.~\ref{fig:bvi-ctc} compares the selection rate statistics on the BVI-DVC \cite{ma2021bvi} and the JVET Common Test Conditions (CTC) \cite{Karczewicz2023ctc}, for luma and chroma, respectively. These statistics have been extracted from ECM-11.0. As can be seen, most tools have roughly stable statistical behaviors in CTC and BVI-DVC, which is a positive sign that they will likely perform as efficiently when deployed on real-world content. In particular, by putting together the stable statistics of MIP from Fig. \ref{fig:versions}, one can roughly conclude that MIP has had reliable matrix training with almost no over-fitting.

\begin{figure}
    \centering
    \includegraphics[width=.5\textwidth]{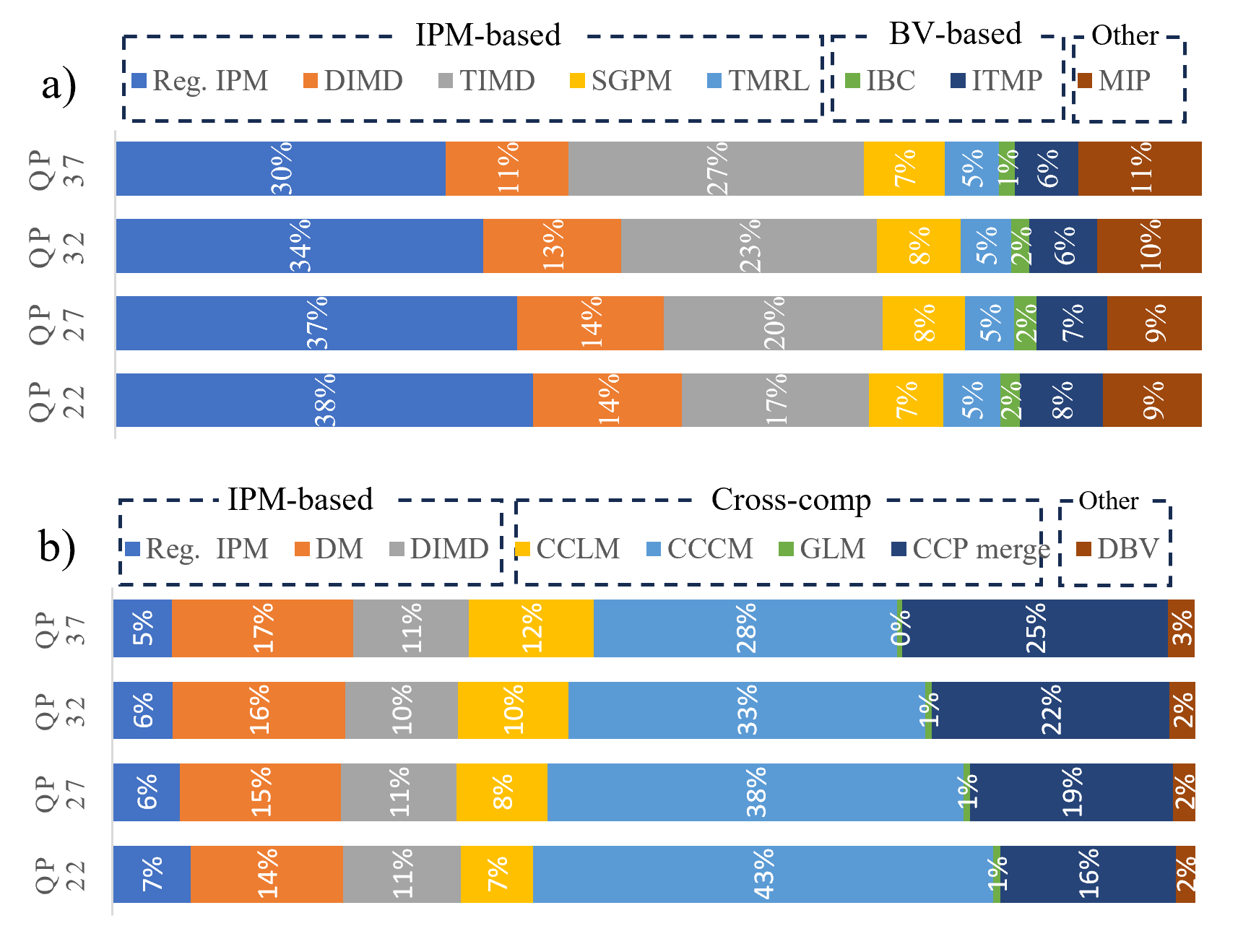}
    \caption{Selection rate of luma (a) and chroma (b) coding tools on different QPs (ECM-11.0).}
    \label{fig:qp}
\end{figure}

Finally, Fig.~\ref{fig:qp} compares the statistics in different Quantization Parameter (QP) values. Similarly, these statistics have been extracted from ECM-11.0. In luma, as the QP increases, the regular IPM is selected less often, while, TIMD gets selected more frequently. This is due to the fact that regular IPM coding requires significantly larger signaling cost compared to decoder-side derivations (\textit{e.g.} TIMD and DIMD), which is not affordable at higher QPs. In chroma, as QP increases, trends can be observed in three tools of CCCM, CCLM and CCP merge. In particular, in higher QPs CCCM gets selected less often, while CCLM and CCP merge are selected more frequently. However, the total selection rate of the four cross-component tools seems to be unchanged. Moreover, the selection rate of chroma DIMD slightly increases in higher QPs.

\section{Conclusion}
\label{sec:conclusion}

This paper presents a quantitative analysis of coding tools that have been developed and adopted in the ECM for the purpose of the post-VVC exploration at JVET. The studied tools particularly concern intra coding luma and chroma of natural video content, with a focus on the selection rate of the coding tools, in terms of percentage of blocks coded with each coding tool. In general, this study shows that simpler tools are being replaced by relatively more advanced tools that require more encoder and/or decoder side processing (\textit{e.g.} template-based coding, texture analysis etc.). Moreover, data-driven tools are trending both in form of offline training (\textit{e.g.} MIP) and online training (\textit{e.g.} cross-component model solver). A trend that will most likely be followed during the standardization phase of JVET. Furthermore, this study did not show a major difference in terms of statistical behaviour when comparing CTC vs. non-CTC content. These observations prove that the exploration phase at the JVET is probably progressing steadily toward the next generation codec. As future work, one might conduct the same quantitative study on inter coding tools of ECM.



\vfill\pagebreak

\balance
\bibliographystyle{IEEEbib}
\bibliography{sample}

\end{document}